\documentclass[aps,prb,twocolumn,floatfix,superscriptaddress]{revtex4}

\usepackage{amsmath}
\usepackage{amssymb}
\usepackage{times}
\usepackage{braket}
\usepackage[pdftex]{graphicx}
\usepackage{color}
\usepackage{blindtext}

\renewcommand{\vec}[1]{\boldsymbol{#1}}

\newcommand{\ed}[1]{{\color{black}{#1}}} 
 
\begin{document}

\title{Topological Hall signatures of magnetic hopfions}

\author{B{\"o}rge G{\"o}bel}
\email[Corresponding author. ]{boerge.goebel@physik.uni-halle.de}
\affiliation{Institut f\"ur Physik, Martin-Luther-Universit\"at Halle-Wittenberg, D-06099 Halle (Saale), Germany}
\affiliation{Max-Planck-Institut f\"ur Mikrostrukturphysik, D-06120 Halle (Saale), Germany}

\author{Collins Ashu Akosa}
\affiliation{RIKEN  Center  for  Emergent  Matter  Science  (CEMS), 2-1  Hirosawa,  Wako,  Saitama  351-0198,  Japan}
\affiliation{Department  of  Theoretical  and  Applied  Physics, African  University  of  Science  and  Technology  (AUST), Km  10  Airport  Road,  Galadimawa,  Abuja  F.C.T,  Nigeria}

\author{Gen Tatara}
\affiliation{RIKEN  Center  for  Emergent  Matter  Science  (CEMS), 2-1  Hirosawa,  Wako,  Saitama  351-0198,  Japan}
\affiliation{RIKEN  Cluster  for  Pioneering  Research  (CPR),2-1  Hirosawa,  Wako,  Saitama,  351-0198  Japan}

\author{Ingrid Mertig}
\affiliation{Institut f\"ur Physik, Martin-Luther-Universit\"at Halle-Wittenberg, D-06099 Halle (Saale), Germany}

\date{\today}

\begin{abstract}
Magnetic hopfions are topologically protected three-dimensional solitons that are constituted by a tube which exhibits a topologically nontrivial spin texture in the cross-section profile and is closed to a torus. Here, we show that the hopfion's locally uncompensated emergent field leads to a topological Hall signature, although the topological Hall effect vanishes on the global level. The topological Hall signature is switchable by magnetic fields or electric currents and occurs independently of the anomalous and conventional Hall effects. It can therefore be exploited to electrically detect hopfions in experiments and even to distinguish them from other textures like skyrmion tubes. Furthermore, it can potentially be utilized in spintronic devices. Exemplarily, we propose a hopfion-based racetrack data storage device and simulate the electrical detection of the hopfions as carriers of information.
\end{abstract}


\maketitle

\section{Introduction}

Over the recent years, non-collinear spin textures have attracted an enormous research interest. Especially the small whirls called `magnetic skyrmions'~\cite{bogdanov1989thermodynamically,muhlbauer2009skyrmion} have outstanding properties which arise from their integer topological charge~\cite{nagaosa2013topological}
\begin{align}
N_\mathrm{Sk}=\frac{1}{4\pi}\int\vec{m}(\vec{r})\cdot\left(\frac{\partial\vec{m}(\vec{r})}{\partial x}\times \frac{\partial\vec{m}(\vec{r})}{\partial y}\right)\,\mathrm{d}^2r
\end{align}
[$\vec{m}(\vec{r})$ is the normalized magnetization density]. In nature, the quasi two-dimensional skyrmions elongate as `strings' or `tubes' along the magnetic field direction. For these textures, the occurrence of an additional contribution to the Hall effect of electrons~\cite{neubauer2009topological,lee2009unusual} has been observed. Due to the accumulation of a Berry phase upon reorientation of their spin with respect to the locally varying texture, the electrons behave as if they would interact with an emergent magnetic field~\cite{nagaosa2013topological,hamamoto2015quantized}
\begin{align}
B_\mathrm{em,\alpha}=\frac{1}{2}\epsilon_{\alpha\beta\gamma}\vec{m}(\vec{r})\cdot\left(\frac{\partial\vec{m}(\vec{r})}{\partial \beta}\times \frac{\partial\vec{m}(\vec{r})}{\partial \gamma}\right).
\end{align}
This effective field can be gigantic compared to experimentally realizable magnetic fields and it points always along the string direction. The topological Hall effect can in principle be used to electrically detect individual magnetic skyrmions that move in between two leads~\cite{maccariello2018electrical,hamamoto2016purely}.

\begin{figure*}[ht!]
  \centering
  \includegraphics[width=\textwidth]{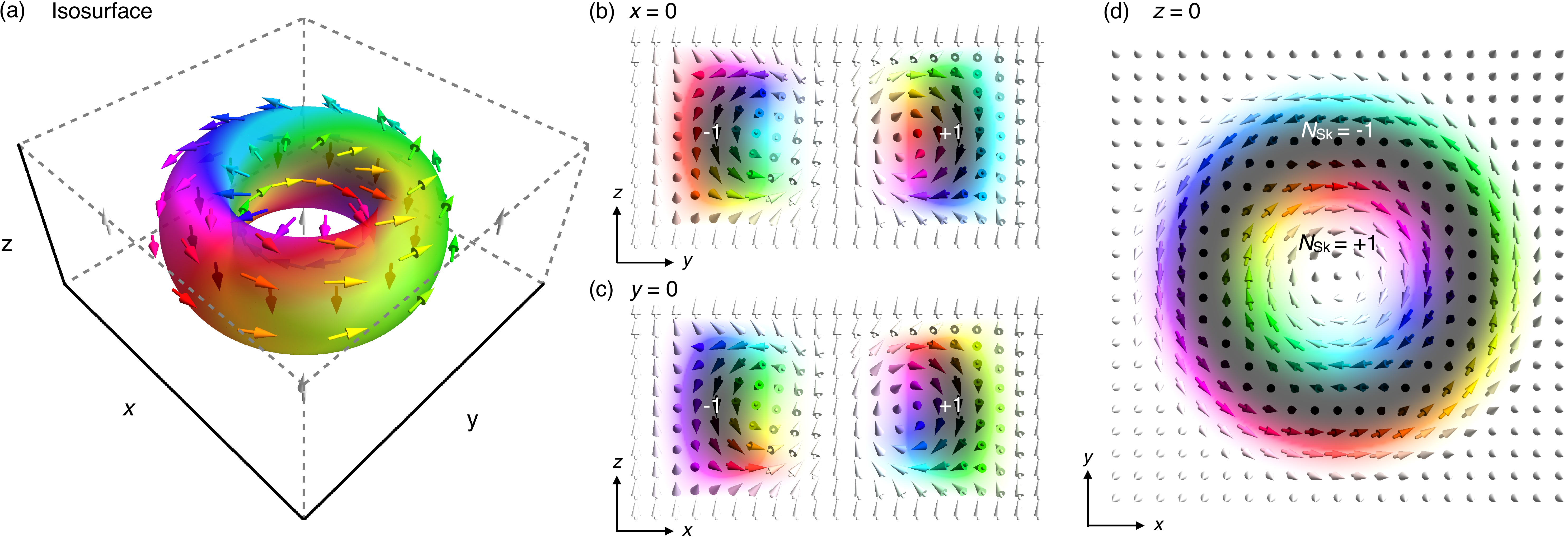}
  \caption{Magnetic hopfion. In (a) the isosurface $m_z=-0.2$ and selected magnetic moments are shown. In (b,c,d) cuts of the texture at $x=0$, $y=0$ and $z=0$ are shown, respectively. The color represents the in-plane spin orientation. White and black represent $+z$ and $-z$ orientations, respectively. The topological charges of the magnetic objects in these two-dimensional cuts are indicated. These quantities can be related with the emergent field perpendicular to the respective cut (cf. Fig.~\ref{fig:emergent}).}
  \label{fig:hopfion}
\end{figure*}

\ed{A related object is the magnetic hopfion~\cite{sutcliffe2007vortex,sutcliffe2018hopfions, liu2018binding,tai2018static,wang2019current,rybakov2019magnetic}. It} can be understood as a skyrmion tube that is closed to a torus [Fig.~\ref{fig:hopfion}(a)] -- this brings about another layer of topological  protection, since a torus cannot be transformed to a string continuously. 
\ed{Recently, magnetic hopfions have been} stabilized in micromagnetic simulations~\cite{liu2018binding,tai2018static,wang2019current}. While considering cross-section textures with higher topological charges or strings that are twisted before forming the closed torus can lead to increased hopf numbers~\cite{whitehead1947expression,wilczek1983linking}
\begin{align}
N_H=-\frac{1}{(4\pi)^2}\int\vec{B}_\mathrm{em}(\vec{r})\cdot\vec{A}(\vec{r})\,\mathrm{d}^3r,
\end{align}
here, we consider the fundamental hopfions (cf. Fig.~\ref{fig:hopfion}) characterized by a hopf number of 1, as stabilized in Refs.~\cite{liu2018binding,tai2018static,wang2019current}. The hopf number is calculated from the texture's emergent field $\vec{B}_\mathrm{em}$ and its corresponding vector potential~\ed{\footnote{The vector potential can be constructed directly from the emergent field in terms of elementary integrals, as shown in Ref.~\cite{rybakov2019magnetic}.}}, fulfilling $\nabla\times\vec{A}=\vec{B}_\mathrm{em}$.

To be precise, in order to be geometrically compatible with the magnetic surrounding, the cross-section profile of the torus is not a conventional skyrmion but an in-plane skyrmion [cf. Figs.~\ref{fig:hopfion}(b,c)], also called `bimeron'~\cite{kharkov2017bound,gobel2018magnetic}. The texture along every cut that includes the $z$ axis resembles the same two bimerons, but due to the deformation of the string upon forming the torus, the magnetization points along different directions [different colors in the two cuts in Figs.~\ref{fig:hopfion}(b,c)]. This is also visible in the $z=0$ cross-section profile in Fig.~\ref{fig:hopfion}(d) that resembles a Bloch-type skyrmionium -- a skyrmion that is positioned in the center of a second skyrmion with mutually reversed spins~\cite{bogdanov1999stability,zhang2018real}. 

In the following, we calculate the topological contribution to the Hall effect of electrons in different measurement setups by means of Landauer-B{\"u}ttiker simulations. We find that hopfions exhibit a local topological Hall signature due to a locally uncompensated emergent field, that cancels only on a global level. Based on these findings, we discuss consequences for the electrical detection and for the spintronic applicability of magnetic hopfions.

\section{Model and methods} 
\begin{figure*}[t]
  \centering
  \includegraphics[width=0.75\textwidth]{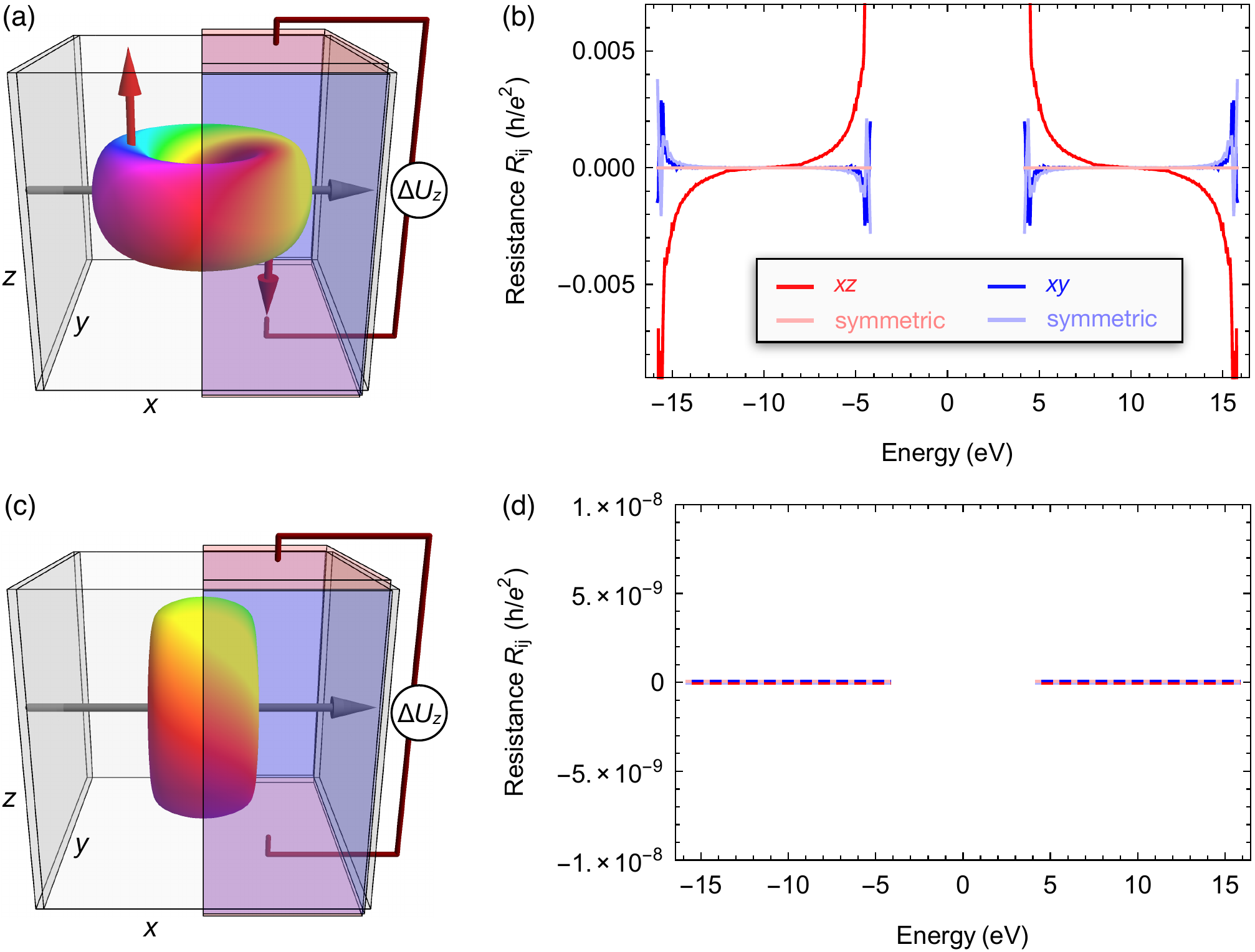}
  \caption{Topological Hall signal using asymmetric leads. In (a) the hopfion is positioned such that the asymmetrically placed detecting leads (red and blue) enclose one vertically-cut half of the hopfion, so that the average emergent field is pointing along $-y$ (cf. arrows in Fig.~\ref{fig:emergent}). The evoked deflection of the current electrons (gray) is indicated by the red arrows. The resulting Hall voltage is detected by the red and blue leads; it gives rise to the corresponding resistance tensor elements shown in (b). The brighter curves show the signal for symmetric leads that cover the whole cube for comparison. In (c,d) the hopfion is reoriented, as indicated.}
  \label{fig:asymmetric}
\end{figure*}

We consider a cuboid-shaped region of a cubic lattice, in which the hopfion is located, and add semi-infinite leads attached to the surfaces of the cuboid, in order to simulate currents and charge accumulations. Depending on the measurement geometry, the six leads either cover the whole cuboid or border only parts of it [cf. blue and red detecting leads in Fig.~\ref{fig:asymmetric}(a,c)] to locally probe the topological Hall signature. 

For simplicity, we consider a single orbital tight-binding model with a hopping term and a Hund's coupling term~\cite{hamamoto2015quantized,gobel2017THEskyrmion,gobel2017QHE}
\begin{align}
 H & = t \sum_{\braket{i,j}} \,c_{i}^\dagger c_{j} + m \sum_{i} \vec{m}_{i} \cdot (c_{i}^\dagger \vec{\sigma}c_{i}).\label{eq:ham}
\end{align}
Here $c_i^\dagger$ and $c_i$ are the spin-dependent creation and annihilation operators of an electron at lattice site $i$. The parameter $t=1\,\mathrm{eV}$ quantifies the nearest-neighbor hopping and $m$ the coupling of electron spins ($\vec{\sigma}$ vector of Pauli matrices) and the magnetic texture $\{\vec{m}_i\}$. If not stated otherwise, we consider the \ed{case close to the} adiabatic limit, $m=10t$, \ed{where spin parallel and anti-parallel states are separated in energy}. The texture for an $N_H=1$ hopfion is taken from Ref.~\cite{sutcliffe2018hopfions}. For a cylindrical region with radius and height L (cylindrical coordinates $\rho=\sqrt{x^2+y^2}$, polar angle $\phi$ and height $z$), the hopfion is described by~\cite{sutcliffe2018hopfions}
\begin{align}
\vec{m}(\vec{r})=\begin{pmatrix} 
\frac{4\Xi\rho(\Omega\cos\phi-(\Lambda-1)\sin\phi)}{(1+\Lambda)^2}\\ 
\frac{4\Xi\rho(\Omega\sin\phi+(\Lambda-1)\cos\phi)}{(1+\Lambda)^2}\\
1-\frac{8\Xi^2\rho^2}{(1+\Lambda)^2}
\end{pmatrix},
\end{align}
with 
\begin{align}
\Omega&=\tan(\pi z/L),\notag\\
\Xi&=(1+(2z/L)^2)\sec(\pi\rho/(2L))/L,\\
\Lambda&=\Xi^2\rho^2+\Omega^2/4.\notag
\end{align}
Outside of this specified region, the magnetization is assumed to point along $\vec{z}$.

In a periodic sample, the band structure of such a single orbital model without considering a texture gives a spin degenerate band that ranges from $-6t$ to $+6t$ in energy. Accordingly, in our system, when the coupling to the texture is considered, the density of states is finite for energies between $-m-6t$ and $+m+6t$. For $m=10t$, the relevant energy range is from $-16\,\mathrm{eV}$ to $16\,\mathrm{eV}$, with a gap between $-4\,\mathrm{eV}$ and $4\,\mathrm{eV}$ due to the considered strong-coupling limit.

To calculate the Hall resistivity, we use a Landauer-B{\"u}ttiker method, as presented in Refs.~\cite{hamamoto2016purely,yin2015topological} for skyrmions. The six leads $i$ are characterized by currents $I_i$ and voltages $U_i$ that are related by the transmission matrix $\underline{T}$
\begin{align}
I_i=\frac{e^2}{h}\sum_jT_{ij}U_j.\label{eq:eqsys}
\end{align} 
For the numerical calculation we use the program package Kwant~\cite{groth2014kwant}.
The Hall resistance $R_{ij}$ and the Hall angle $\theta_{ij}$ are given by 
\begin{align}
R_{ij}=\frac{\Delta U_j}{I_i},\quad\quad
\theta_{ij}=\frac{\Delta U_j}{\Delta U_i},
\end{align}
respectively, with $i,j=x,y,z$ \ed{and $i\neq j$}.

In general, each Hall resistance tensor element $R_{ij}$ is determined by several contributions. For skyrmionic textures, one typically considers the conventional Hall effect (HE), the anomalous Hall effect (AHE) and the topological Hall effect (THE)~\cite{neubauer2009topological,lee2009unusual} of electrons. The conventional Hall effect is proportional to an externally applied magnetic field~\cite{hall1879new} along the perpendicular direction, and the anomalous Hall effect~\cite{hall1881xviii} arises due to spin-orbit coupling (SOC) and (typically~\ed{\footnote{There are recent reports that also coplanar textures without a net magnetization can exhibit an anomalous Hall effect~\cite{chen2014anomalous,nakatsuji2015large,nayak2016large}.}}) due to a net magnetization~\cite{nagaosa2010anomalous}. The topological Hall resistance is proportional to the average emergent field $\braket{B_\mathrm{em}}$ along the perpendicular direction. This effect arises purely due to the existence of a topologically non-trivial spin texture~\cite{bruno2004topological}, even in the absence of SOC. In order to isolate this fundamental contribution, we do not consider SOC nor an external magnetic field. 

\section{Results and discussion}
\subsection{Vanishing topological Hall effect on a global level}
First, we consider a cube of size $(2L+1)\times (2L+1)\times (2L+1)$ sites with a hopfion in its center (here modeled by $L=10$). Six leads border the whole cube to simulate a symmetric measurement. 
We model an applied current along the $x$ direction by adding a small bias voltage $V_{-x}=-V_{+x}=1\,\mathrm{mV}$ and are interested in the charge accumulation at the other terminals. Therefore, the currents were set to zero $I_{-y}=I_{+y}=I_{-z}=I_{+z}=0$. Solving Eqs.~\eqref{eq:eqsys} gives $R_{xz}=0$ for this geometry~\footnote{The calculated values are within the order of magnitude of the numerical precision.}; cf. light red curve in Figs.~\ref{fig:asymmetric}(b).

\begin{figure}[t]
  \centering
  \includegraphics[width=\columnwidth]{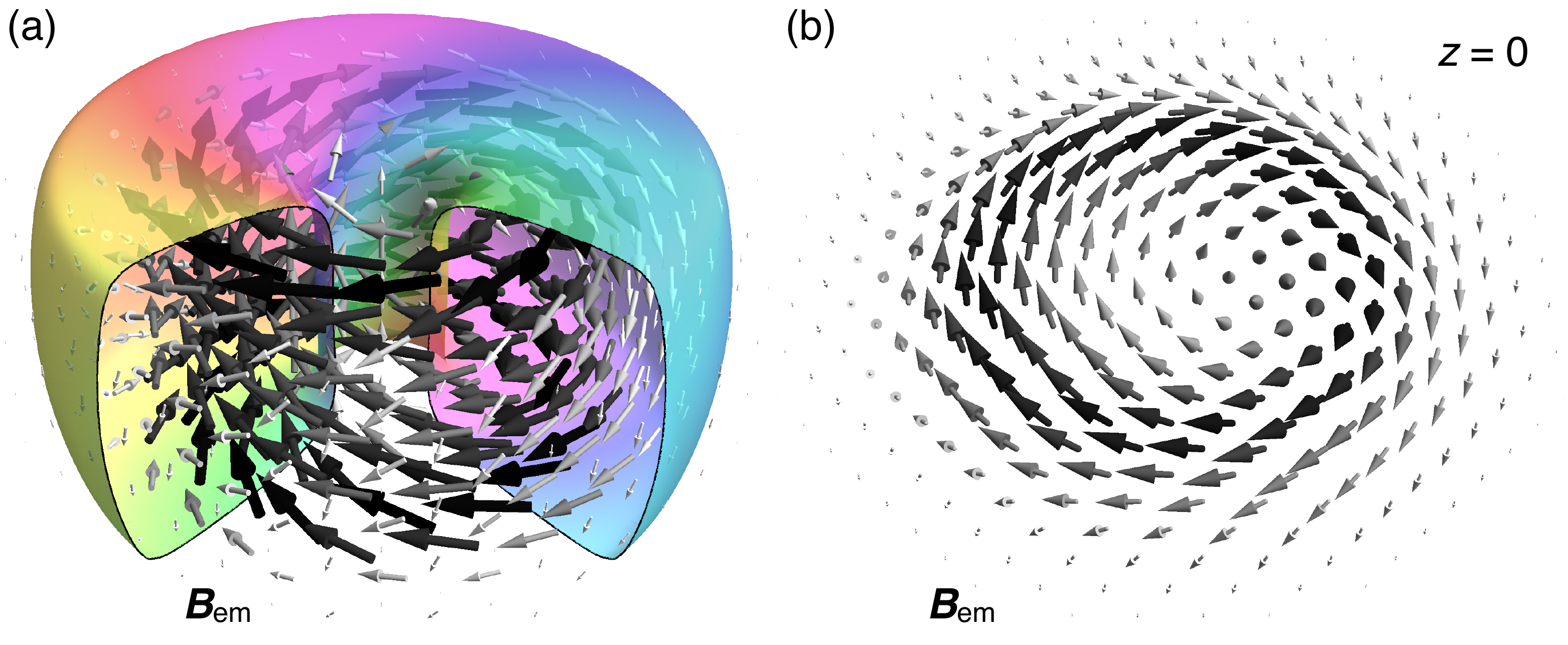}
  \caption{Emergent field. (a) The arrows visualize the emergent field of a hopfion, where the gray-level represents the magnitude. The colored shell shows a part of the isosurface for $m_z=0.75$ for the purpose of better visibility. In (b) a cut at $z=0$ is shown.}
  \label{fig:emergent}
\end{figure}

This observation can be understood by considering the emergent field of a hopfion (Fig.~\ref{fig:emergent}). For the cuts at $x=0$ and $y=0$, the spin texture resembles two in-plane skyrmions [Figs.~\ref{fig:hopfion}(b,c)]. Consequently, the emergent field points along the perpendicular direction, i.\,e. along the torus. For the cut at $z=0$ [Figs.~\ref{fig:hopfion}(d)], the profile is a skyrmionium with a positive out-of-plane emergent field in the center and a negative field for the outer ring. Consequently, the emergent field forms a whirl-like vector field with positive $z$ components in the center and opposite orientations near the hopfion's edge [cf. Fig.~\ref{fig:emergent}(b) for better visibility]. The emergent field is characterized by a finite toroidal moment $\vec{t}\propto\int \vec{r}\times\vec{B}_\mathrm{em}(\vec{r})\,\mathrm{d}^3r$. The field vanishes on average, and even on a discrete lattice its average in-plane component vanishes perfectly by symmetry explaining $R_{xz}=0$.

The other resistance tensor element, $R_{xy}$, is negligible for most energies as well [light blue curve in Fig.~\ref{fig:asymmetric}(b)]. However, finite values occur close to the band edge which originate from the discrete lattice and the small hopfion size~\footnote{A large angle between two neighbored magnetic moments effectively leads to a decrease of the hopping amplitude. Consequently, for energies near the band edge, the electrons are located predominantly in areas with smaller angles between neighbored spins~\cite{hamamoto2016purely}. The electrons effectively feel a small emergent field which is oriented out of the hopfion plane and causes a finite $R_{xy}$ signal.}. All in all, a hopfion does not exhibit a considerable Hall response in a global measurement.

\subsection{Local topological Hall signature}
Even though a globally vanishing emergent field leads to a compensated signal for symmetrically placed leads, the emergent field does not vanish locally. For this reason, in a second simulation, we apply asymmetric leads with respect to the hopfions's center. As shown in Figs.~\ref{fig:asymmetric}(a,c), the leads parallel to the $xy$ and $xz$ planes are only $L$ sites wide and cover only one half of the cube. Consequently, the emergent field in the volume between those leads predominantly determines the magnitude of the topological Hall resistance. Here, the field points along $-\vec{y}$ on average.

This time, a pronounced $R_{xz}$ signature is observed [red curve in Fig.~\ref{fig:asymmetric}(b)], in agreement with the phenomenological Lorentz-force argument: the force points along $-z$ for an average emergent field along $-y$ and a current along $x$. Consequently, the applied current (gray) is deflected along $-\vec{z}$ (red arrow). The hopfion exhibits a considerable local topological Hall effect, with Hall angles of up to $6\%$ (shown in Fig.~S1 in the Supplemental Material~\cite{SupplementalMaterial}). Likewise, if the contacts are attached at the other half of the cube, the sign of $R_{xz}$ is reversed -- both contributions cancel in a global measurement.

The sign of the $R_{xz}$ signal is energy dependent and determined by the carrier character and the local spin alignment with respect to the texture. Therefore, it changes comparing the positive and negative energy states (parallel vs. anti-parallel spin alignment), and at the energies $\pm m = \pm 10\,\mathrm{eV}$, where the predominant carrier concentrations changes from electronic to hole-like, similar to Ref.~\cite{gobel2017THEskyrmion}. 

In another simulation, we rotated the hopfion as shown in Fig.~\ref{fig:asymmetric}(c)]. In this configuration, the relevant emergent field vanishes on average, which is why the numerical simulations return vanishing Hall coefficients [Fig.~\ref{fig:asymmetric}(d)]. 

Following from our results, hopfions can even be distinguished from skyrmion strings in a sample which is magnetized along $\vec{z}$: For skyrmions, the Hall response is limited to the $xy$ resistance tensor element. The $xz$ element is zero since the emergent field of a skyrmion string points always along $z$.
Another noteworthy property that can be concluded from these results is that for a hopfion the predominant topological Hall voltage arises parallel to the net magnetization and the stabilizing magnetic field; the $R_{xz}$ tensor element is non-zero. Therefore, if a current is applied within the hopfion plane, as in Fig.~\ref{fig:asymmetric}(a), this particular locally detected topological Hall signal is not superimposed by the anomalous Hall effect and the conventional Hall effect that both have only finite $xy$ elements. This makes hopfions unique and allows for a pure detection of the topological Hall effect.

The above findings are not restricted to the adiabatic limit. A local topological Hall signature arises even for a weak coupling of electron spins and the magnetic texture ($m=2/3\,t$ in Fig.~S2 in the Supplemental Material~\cite{SupplementalMaterial}), even though the validity of the emergent field interpretation is limited in that case.

%


\subsection{Hopfion-based racetrack storage device}
\begin{figure}
  \centering
  \includegraphics[width=\columnwidth]{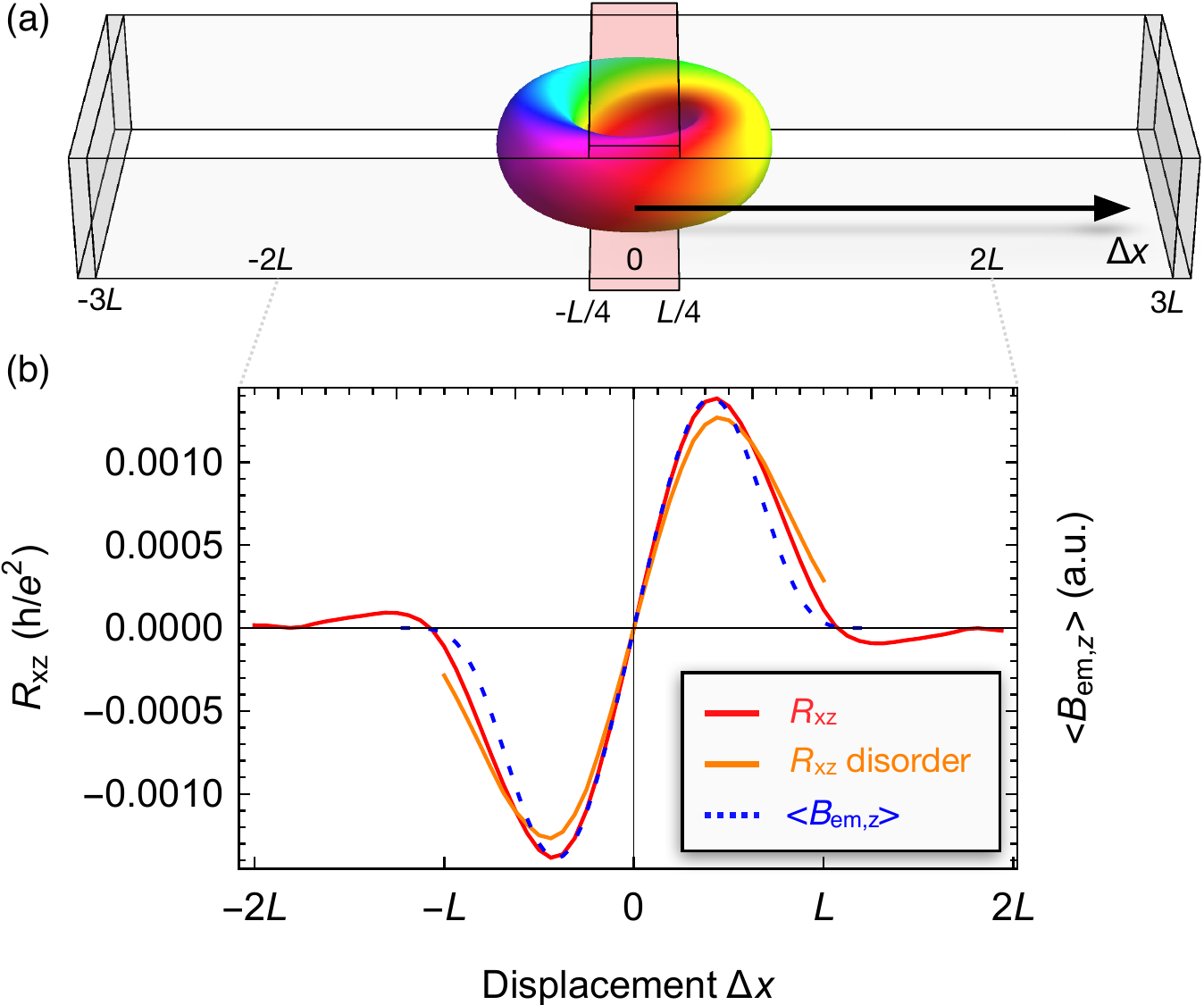}
  \caption{Local detection of hopfions in a nano-stripe. In (a) the considered geometry is shown with the displacement $\Delta x$ of the hopfion with respect to the detecting leads, as indicated. A current, flowing between the gray contacts (along $x$), leads to a Hall voltage between the two red contacts (along $\pm$ z). Panel (b) shows the calculated Hall resistance signal (red). For the orange curve, disorder has been taken into account (see main text). The average emergent field between the two red leads is shown as a blue, dashed curve. The simulated racetrack has the dimension: $(6L+1)\times (2L+1)\times (L+1)$ sites, here $L=16$ sites, which allows to simulate displacements between $-2L$ and $+2L$. The red leads have a width of $9$ sites and the Fermi energy is $E_F=-14.0\,\mathrm{eV}$.}
  \label{fig:racetrack}
\end{figure}

A potential spintronic device that can exploit the locally occurring topological Hall effect is a hopfion-based racetrack data storage. Similar to the initially proposed racetracks based on domain walls~\cite{parkin2004shiftable,parkin2008magnetic,parkin2015memory} or the later proposed racetracks based on skyrmions~\cite{sampaio2013nucleation}, a hopfion-based racetrack is a nano-stripe, where the bits of information are encoded by the presence or absence of hopfions at specific positions. Therefore, the hopfions need to be written, deleted, moved and read. Recently, the current-driven motion of hopfions by spin-transfer and spin-orbit torques has been simulated~\cite{wang2019current}. The hopfions, that are oriented in the $xy$ plane, move along the track, without any transverse deflection, like in Fig.~\ref{fig:racetrack}(a). This is highly desirable for a racetrack storage device and can be considered a great advantage for the applicability of hopfions compared to skyrmions. Our finding provides a method to detect the bits in such a geometry by adding detecting leads, similar to the case of skyrmions presented in Refs.~\cite{maccariello2018electrical,hamamoto2016purely}; cf. Fig.~\ref{fig:racetrack}(a). However, here, the leads are positioned perpendicular to $z$ (red). 

As expected from our prior findings, the calculated signal $R_{xz}$ [Fig.~\ref{fig:racetrack}(b)] is antisymmetric with respect to the hopfion displacement $\Delta x=0$ and exhibits two significant peaks when the hopfion is displaced by approximately $\Delta x = \pm 0.4L$. In these cases, the space-dependent emergent field between the leads 
\begin{align}
\braket{\vec{B}_\mathrm{em}}=\int_{x=-L/4}^{L/4}\int_{y=-L}^L\int_{z=-L/2}^{L/2}\vec{B}_\mathrm{em}(\vec{r}-\Delta \vec{r})\,\mathrm{d}\vec{r}\notag
\end{align}
(blue dashed curve) has its extremum as well. The relevant emergent field points along $\pm \vec{y}$, respectively. The good agreement of both curves implies that the Hall resistance is proportional to the relevant emergent field, like in Ref.~\cite{goebel2019electrical}. 

For a more realistic simulation, we considered also disorder by adding site-dependent random on-site energies that range between $-t/2$ and $+t/2$~\cite{ndiaye2017topological}, which corresponds to a mean free path of 128 lattice constants. An average over 600 impurity configurations has been taken. The calculated signal still features two antisymmetric peaks, even though disorder decreases their magnitudes [cf. orange curve in Fig.~\ref{fig:racetrack}(b)]. The $R_{xz}$ signal is robust compared to the finite-size mediated $R_{xy}$ that decreases by more than $60\%$ when disorder is taken into account (cf. Fig.~S3 in the Supplemental Material~\cite{SupplementalMaterial}).

\section{Conclusion}
In summary, we have shown that magnetic hopfions exhibit a local topological Hall signature which is compensated on a global level. As long as the applied current is not perpendicular to the hopfion plane, the toroidal emergent field of the hopfion leads to a local deflection of electrons along the out-of-plane direction. This direction coincides with the net magnetization of a hopfion host and the stabilizing magnetic field. Consequently, this topological Hall signature is expected to be detectable in a pure manner, i.\,e., without anomalous or conventional Hall effects. It allows to detect hopfions in experiments and to distinguish them from skyrmion strings.

This fundamental finding can be utilized in spintronic devices. Exemplarily, we have discussed a racetrack storage device. Here, the hopfion is a highly promising carrier of information, since its globally compensated emergent field leads to the absence of a skyrmion Hall effect. Therefore, our work does not only contribute to the fundamental understanding of three-dimensional magnetic solitons but opens also avenues towards the realization of innately three-dimensional spintronic applications.

Furthermore, we have shown that the local topological Hall signature depends on the hopfion's orientation, which in micromagnetic simulations is controlled by the stabilizing magnetic field. For this reason, electrical currents can be controlled by tilting an external field. Conversely, if hopfions were found to be stable without magnetic fields in different samples, our results imply that the application of an electrical current would lead to a reorientation of the hopfion: By analogy with skyrmionic systems, a topological Hall effect is accompanied by a torque that manipulates the texture itself. Without a stabilizing field, the application of a current in the setup of Fig.~\ref{fig:asymmetric}(a) would rotate the hopfion to the state in Fig.~\ref{fig:asymmetric}(c), since the torque would have opposite signs for the two sides of the hopfion~\footnote{By analogy with the local topological Hall effect of electrons, the two sides of the hopfion would move oppositely to the red arrows in Fig.~\ref{fig:asymmetric}(a).}. As long as a considerable stabilizing field is present, this effect is however suppressed~\footnote{For example, in the simulations of Ref.~\cite{wang2019current} the effect was not observed. Furthermore, even for very small stabilizing fields for which the competition of forces could lead to an intermediate orientation of the hopfion between Figs.~\ref{fig:asymmetric}(a) and (c), the hopfion would rotate back to the initial configuration once the current is tuned off again.}.\\

\begin{acknowledgments}
This work is supported by TRR 227 of Deutsche Forschungsgemeinschaft (DFG). C. A. A. and G. T. acknowledge support by Grant-in-Aid for Exploratory Research (No. 16K13853) and Grant-in-Aid for Scientific Research (B) (No. 17H02929) from Japan Society for the Promotion of Science (JSPS). 
\end{acknowledgments}

\bibliography{short,MyLibrary}
\bibliographystyle{apsrev}

\end{document}